\documentclass[aps,prd,eqsecnum,preprint,floats,epsf,superscriptaddress,nofootinbib,letter]{revtex4}

\usepackage{epsfig,color}
\usepackage{amssymb,amsmath}

\usepackage[colorlinks,
            linkcolor=black,
            anchorcolor=black,
            citecolor=black
            ]{hyperref}

\def\be{\begin{eqnarray}}
\def\en{\end{eqnarray}}
\def\non{\nonumber}
\def\ov{\overline}

\def\B{{\cal B}}
\def\S{{\cal S}}

\def\T{{\cal T}}

\def\X{{\cal X}}

\def\lsim{ {\ \lower-1.2pt\vbox{\hbox{\rlap{$<$}\lower5pt\vbox{\hbox{$\sim$}
}}}\ } }
\def\gsim{ {\ \lower-1.2pt\vbox{\hbox{\rlap{$>$}\lower5pt\vbox{\hbox{$\sim$}
}}}\ } }

\begin{document}

\begin{flushright}
May, 2017 \\ NCTS-PH/1708
\end{flushright}

\vspace{0.0cm}

\begin{center}
{\large\bf Quantum Numbers of $\Omega_c$ States and Other Charmed Baryons}

\vskip 0.5cm

Hai-Yang Cheng$^{a}$
~and Cheng-Wei Chiang$^{b,a,c}$

\vskip 0.3cm

{\it $^a$ Institute of Physics, Academia Sinica, Taipei 11529, Taiwan, Republic of China}
\\
{\it $^b$ Department of Physics, National Taiwan University, Taipei 10617, \\
Taiwan, Republic of China}
\\
{\it $^c$ Physics Division, National Center for Theoretical Sciences, Hsinchu 30013, \\
Taiwan, Republic of China}

\vskip 0.5cm

\date{\today}

{\large Abstract}
\end{center}
Possible spin-parity quantum numbers for excited charmed baryon resonances are discussed in this work. Our main results are:
(i)
Among the five newly observed $\Omega_c$ states, we have identified $\Omega_c(3090)$ and $\Omega_c(3119)$ with radially excited $\frac12^+(2S)$ and $\frac32^+(2S)$ states, respectively, and $\Omega_c(3000)$ with $\frac12^-(1P)$ and $S=\frac32$. The two states $\Omega_c(3050)$ and $\Omega_c(3066)$ form a $P$-wave $(\frac32^-,\frac52^-)$ doublet.
(ii) The widths of $\Omega_c(3066)$ and $\Xi'_c(2930)$ are calculable within the framework of heavy hadron chiral perturbation theory.
(iii) Since the width of $\Omega_{c0}(\frac12^-)$ is of order 410 MeV, not all observed narrow $\Omega_c$ baryons can be identified with $1P$ states.
(iv)
For the antitriplet $\Lambda_c$ and $\Xi_c$ states, their Regge trajectories for the orbital excitations of $\frac12^-$ and $\frac32^-$ are parallel to each other. Based on this nice property of parallelism, we see that the highest state $\Lambda_c(2940)$ does not fit if its quantum numbers are $\frac32^-$ as found by LHCb.  We suggest that $\Lambda_c(2940)^+$ is most likely the $\frac12^-(2P)$ state.
(v)
The charmed baryon $\Sigma_c(2800)$ cannot be a $\frac12^-$ state; otherwise, its width will be over 400~MeV, too large compared to the measured one.
(vi)
In the study of Regge trajectories of $\Xi'_c$ states, we find a missing state. It should have quantum numbers $\frac52^-$ with a mass around 2920~MeV.
(vii)
Antitriplet and sextet states are classified according to their $J^P(nL)$ quantum numbers. The mass differences between $\Xi_c$ and $\Lambda_c$ in the antitriplet states clearly lie between 180 and 200~MeV. Moreover, the mass splitting  between $\Omega_c$ and $\Xi'_c$ is found to be very close to the one between $\Xi'_c$ and $\Sigma_c$ for five different sets of sextet multiplets. This lends a strong support to the quantum number assignment to the sextet states in this work.

\newpage

\section{Introduction}

\begingroup
\squeezetable
\begin{table}[!]
\caption{\scriptsize Mass spectra and widths (in units of MeV) of the observed charmed baryons. Experimental values are taken from the Particle Data Group~\cite{PDG}.  For the masses and widths with a superscript $^\dagger$ and $^*$, we have taken into account the recent measurements of LHCb~\cite{LHCb:Lambdac2880} or Belle~\cite{Belle:Xic}, respectively, for a weighted average. For $\Xi_c(3055)^0$, we quote the result from Belle~\cite{Belle:Kato}.  For the five new $\Omega_c$ states, we quote~\cite{LHCb:Omegac}.
}
\label{tab:spectrum}
\begin{ruledtabular}
\begin{tabular}{c c  c c c }
~~~~~~State~~~~~~~ & ~~$J^P$~~ &
~~~~~~~~~Mass~~~~~~~~~ & ~~~~Width~~~~ &~Decay modes~\\
\hline
 $\Lambda_c^+$ & ${1\over 2}^+$  & $2286.46\pm0.14$ & & weak  \\
 $\Lambda_c(2595)^+$ & ${1\over 2}^-$  & $2592.25\pm0.28$ &
 $2.6\pm0.6$ & $\Lambda_c\pi\pi,\Sigma_c\pi$ \\
 $\Lambda_c(2625)^+$ & ${3\over 2}^-$ & $2628.11\pm0.19$ &
 $<0.97$ & $\Lambda_c\pi\pi,\Sigma_c^{(*)}\pi$ \\
 $\Lambda_c(2765)^+$ & $?^?$  & $2766.6\pm2.4$ & $50$ & $\Sigma_c\pi,\Lambda_c\pi\pi$ \\
 $\Lambda_c(2860)^+$ & ${3\over 2}^+$ & ${2856.1^{+2.3}_{-5.9}}^\dagger$ & ${67.6^{+11.8}_{-21.6}}^\dagger$
 & $\Sigma_c^{(*)}\pi,D^0p,D^+n$ \\
 $\Lambda_c(2880)^+$ & ${5\over 2}^+$ & $2881.64\pm0.25^\dagger$ & $5.6\pm0.7^\dagger$
 & $\Sigma_c^{(*)}\pi,\Lambda_c\pi\pi,D^0p,D^+n$ \\
 $\Lambda_c(2940)^+$ & $?^?$  & $2939.8\pm1.4^\dagger$ & $20\pm6^\dagger$ &
 $\Sigma_c^{(*)}\pi,\Lambda_c\pi\pi,D^0p,D^+n$ \\
 $\Sigma_c(2455)^{++}$ & ${1\over 2}^+$ & $2453.97\pm0.14$ &
 $1.89^{+0.09}_{-0.18}$ & $\Lambda_c\pi$ \\
 $\Sigma_c(2455)^{+}$ & ${1\over 2}^+$ & $2452.9\pm0.4$ &
 $<4.6$ & $\Lambda_c\pi$\\
 $\Sigma_c(2455)^{0}$ & ${1\over 2}^+$ & $2453.75\pm0.14$
 & $1.83^{+0.11}_{-0.19}$ & $\Lambda_c\pi$ \\
 $\Sigma_c(2520)^{++}$ & ${3\over 2}^+$  & $2518.41^{+0.21}_{-0.19}$
 & $14.78^{+0.30}_{-0.40}$ & $\Lambda_c\pi$\\
 $\Sigma_c(2520)^{+}$ & ${3\over 2}^+$  & $2517.5\pm2.3$
 & $<17$ & $\Lambda_c\pi$ \\
 $\Sigma_c(2520)^{0}$ & ${3\over 2}^+$ & $2518.48\pm0.20$
 & $15.3^{+0.4}_{-0.5}$ & $\Lambda_c\pi$ \\
 $\Sigma_c(2800)^{++}$ & $?^?$  & $2801^{+4}_{-6}$ & $75^{+22}_{-17}$ &
 $\Lambda_c\pi,\Sigma_c^{(*)}\pi,\Lambda_c\pi\pi$ \\
 $\Sigma_c(2800)^{+}$ & $?^?$  & $2792^{+14}_{-~5}$ & $62^{+64}_{-44}$ &
 $\Lambda_c\pi,\Sigma_c^{(*)}\pi,\Lambda_c\pi\pi$ \\
 $\Sigma_c(2800)^{0}$ & $?^?$  & $2806^{+5}_{-7}$ & $72^{+22}_{-15}$ &
 $\Lambda_c\pi,\Sigma_c^{(*)}\pi,\Lambda_c\pi\pi$\\
 $\Xi_c^+$ & ${1\over 2}^+$  & $2467.93^{+0.28}_{-0.40}$ & & weak \\
 $\Xi_c^0$ & ${1\over 2}^+$  & $2470.85^{+0.28}_{-0.40}$ & & weak \\
 $\Xi'^+_c$ & ${1\over 2}^+$  & $2578.3\pm0.5^*$ & & $\Xi_c\gamma$ \\
 $\Xi'^0_c$ & ${1\over 2}^+$  & $2579.2\pm0.5^*$ & & $\Xi_c\gamma$ \\
 $\Xi_c(2645)^+$ & ${3\over 2}^+$  & $2645.7\pm0.3^*$ & $2.1\pm0.2^*$ & $\Xi_c\pi$ \\
 $\Xi_c(2645)^0$ & ${3\over 2}^+$  & $2646.3\pm0.3^*$ & $2.35\pm0.22^*$ & $\Xi_c\pi$ \\
 $\Xi_c(2790)^+$ & ${1\over 2}^-$  & $2791.5\pm0.6^*$ & $8.9\pm1.0^*$ & $\Xi'_c\pi,\Xi_c\pi,\Lambda_c\ov K$ \\
 $\Xi_c(2790)^0$ & ${1\over 2}^-$  & $2794.8\pm0.6^*$ & $10.0\pm1.1^*$ &  $\Xi'_c\pi,\Xi_c\pi,\Lambda_c\ov K$\\
 $\Xi_c(2815)^+$ & ${3\over 2}^-$  & $2816.7\pm0.3^*$ & $2.43\pm0.26^*$ & $\Xi^*_c\pi,\Xi_c\pi\pi,\Xi_c'\pi$ \\
 $\Xi_c(2815)^0$ & ${3\over 2}^-$  & $2820.2\pm0.3^*$ & $2.54\pm0.25^*$ & $\Xi^*_c\pi,\Xi_c\pi\pi,\Xi_c'\pi$ \\
$\Xi_c(2930)^0$ & $?^?$  & $2931\pm6$ & $36\pm13$
 & $\Lambda_c \ov K,\Sigma_c\ov K,\Xi_c\pi,\Xi'_c\pi$ \\
 $\Xi_c(2970)^+$ & $?^?$  & $2966.7\pm0.8^*$ & $24.6\pm2.0^*$
 & $\Sigma_c \ov K,\Lambda_c \ov K\pi,\Xi_c\pi\pi$  \\
 $\Xi_c(2970)^0$ & $?^?$  & $2970.6\pm0.8^*$ & $29\pm3^*$
 & ~$\Sigma_c \ov K,\Lambda_c \ov K\pi,\Xi_c\pi\pi$~ \\
 $\Xi_c(3055)^+$ & $?^?$ & $3055.1\pm1.7$ & $11\pm4$ &
 $\Sigma_c \ov K,\Lambda_c \ov K\pi,D\Lambda$  \\
 $\Xi_c(3055)^0$ & $?^?$  & $3059.0\pm0.8$ & $6.4\pm2.4$ &
 $\Sigma_c \ov K,\Lambda_c \ov K\pi,D\Lambda$  \\
 $\Xi_c(3080)^+$ & $?^?$  & $3076.94\pm0.28$ & $4.3\pm1.5$ &
 $\Sigma_c \ov K,\Lambda_c \ov K\pi,D\Lambda$  \\
 $\Xi_c(3080)^0$ & $?^?$  & $3079.9\pm1.4$ & $5.6\pm2.2$
 & $\Sigma_c \ov K,\Lambda_c \ov K\pi,D\Lambda$ \\
$\Xi_c(3123)^+$ & $?^?$  & $3122.9\pm1.3$ & $4.4\pm3.8$
 & $\Sigma_c^* \ov K,\Lambda_c \ov K\pi,D\Lambda$ \\
 $\Omega_c^0$ & ${1\over 2}^+$  & $2695.2\pm1.7$ & & weak \\
 $\Omega_c(2770)^0$ & ${3\over 2}^+$ & $2765.9\pm2.0$ & & $\Omega_c\gamma$ \\
 $\Omega_c(3000)^0$ & $?^?$ & $3000.4^{+0.4}_{-0.5}$ & $4.5\pm0.7$ & $\Xi_c\ov K$ \\
 $\Omega_c(3050)^0$ & $?^?$ & $3050.2^{+0.3}_{-0.5}$ & $0.8\pm0.2$ & $\Xi_c\ov K$ \\
 $\Omega_c(3066)^0$ & $?^?$ & $3065.6^{+0.4}_{-0.6}$ & $3.5\pm0.4$ & $\Xi_c\ov K$ \\
 $\Omega_c(3090)^0$ & $?^?$ & $3090.2^{+0.7}_{-0.8}$ & $8.7\pm1.3$ & $\Xi_c^{(\prime)}\ov K$ \\
 $\Omega_c(3119)^0$ & $?^?$ & $3119.1^{+1.0}_{-1.1}$ & $1.1\pm0.9$ & $\Xi_c^{(\prime)}\ov K$ \\
\end{tabular}
\end{ruledtabular}
\end{table}
\endgroup

Charmed baryon spectroscopy provides an ideal place for studying
the dynamics of the light quarks in the environment of a heavy
quark. The observed mass spectra and decay widths of singly charmed baryons are
summarized in Table \ref{tab:spectrum}.
By now, the $J^P={1\over 2}^+,\frac12^-,\frac32^+,\frac32^-$ and $\frac52^+$ antitriplet states $\Lambda_c,\Xi_c$ and
$J^P={1\over 2}^+,{3\over 2}^+$ sextet states $\Omega_c,\Xi'_c,\Sigma_c$
are  established (see Table \ref{tab:3and6} below for details). Notice that except for the parity of the lightest
$\Lambda_c^+$ and the heavier ones $\Lambda_c(2880)^+$ \cite{Belle:Lamc2880,LHCb:Lambdac2880} and $\Lambda_c(2860)^+$ \cite{LHCb:Lambdac2880}, none of the other $J^P$ quantum numbers given in
Table \ref{tab:spectrum} have been measured. One has to rely on the
quark model to determine the $J^P$ assignments.

For a long time,
only two ground states had been observed thus far for the $\Omega_c$ baryons: $\frac12^+$ $\Omega_c^0$ and $\frac32^+$ $\Omega_c(2770)^0$. The latter was seen by BaBar in the electromagnetic decay
$\Omega_c(2770)\to\Omega_c\gamma$ \cite{BaBar:Omegacst}. The mass difference between $\Omega_c^*$ and $\Omega_c$ is too small for any strong decay to occur.
Very recently, LHCb has explored this sector and observed five new, narrow excited $\Omega_c$ states decaying into $\Xi_c^+K^-$: $\Omega_c(3000)$, $\Omega_c(3050)$, $\Omega_c(3066)$, $\Omega_c(3090)$ and $\Omega_c(3119)$ \cite{LHCb:Omegac}. This has triggered a lot of interest in attempting to identify their spin-parity quantum numbers \cite{Agaev,Chen:Pwave,Karliner:2017,Yang,Zhao,Wangwei,Huang,Wangzg,Padmanath,Chen:Omegac,Zhao:Omegac,Aliev,Kim,Agaev_2}.

In this work we shall use the predictions of the heavy quark-light diquark model and the Regge trajectories in conjunction with other model calculations to study the spin-parity quantum numbers of sextet and antitriplet charmed baryons, especially the newly discovered $\Omega_c$ resonances.

\section{Spectroscopy}
The charmed baryon spectroscopy has been studied extensively in
various models.  It appears that  the spectroscopy is well described by the model based on the relativitsic heavy quark-light diquark model advocated by Ebert, Faustov and Galkin (EFG) \cite{Ebert:2011} (see also \cite{Chen:2014}). Indeed, the
quantum numbers $J^P=\frac52^+$ of $\Lambda_c(2880)$ have been correctly predicted in the model based on the diquark idea \cite{Selem} even before its discovery in the Belle experiment~\cite{Belle:Lamc2880}.
Based on the heavy quark-light diquark model, EFG have constructed the Regge trajectories of heavy baryons for orbital and radial excitations; all available experimental data on heavy baryons fit nicely to linear Regge trajectories, namely, the trajectories in the $(J,M^2)$ and $(n_r,M^2)$ planes for orbitally and radially excited heavy baryons, respectively:
\begin{eqnarray}
J=\alpha M^2+\alpha_0, \qquad n_r=\beta M^2+\beta_0,
\end{eqnarray}
where $J$ is the baryon spin, $M$ is the baryon mass, $n_r$ is the radial excitation quantum number, $\alpha$, $\beta$ are the slopes and $\alpha_0$, $\beta_0$ are the intercepts. The Regge trajectories can be plotted for charmed baryons with natural $(P=(-1)^{J-1/2})$ and unnatural $(P=(-1)^{J+1/2})$ parities.
We have proposed in \cite{Cheng:2015}  to employ the predictions of the spin-parity quantum numbers of charmed baryons and their masses in \cite{Ebert:2011} as a theoretical benchmark, where the linearity, parallelism and equidistance of the Regge trajectories were verified in their calculations.

\subsection{$\Omega_c$ states}

\begin{table}[t]
\caption{Mass spectrum of the $\Omega_c$ states. 
Numbers inside the parentheses are our suggested assignments  for the masses of the newly observed $\Omega_c$ states. The subscripts $l$ and $h$ denote light and heavy states, respectively, as explained in the text.}
\label{tab:omegac}
\begin{ruledtabular}
\begin{tabular}{c ccc c c}
State & Ebert et al. & Shah et al. & Chen et al. & Agaev et al. & Expt. \\
~~~$nL,J^P$~~~ & \cite{Ebert:2011} & \cite{Shah} &  \cite{Chiu} & \cite{Agaev} &  \cite{LHCb:Omegac} \\ \hline
$1S,1/2^+$ & 2698 & 2695 & $2695\pm24\pm15$ & $ 2685\pm123$ & $2695.2\pm2.0$ \\
$2S,1/2^+$ & 3088 & 3100 & & $3066\pm138$  &  (3090) \\
$1S,3/2^+$ & 2768 & 2767 & $2781\pm12\pm22$  & $2769\pm89$ & $2765.9\pm2.0$ \\
$2S,3/2^+$ & 3123 & 3126 & $$   & $3119\pm114$  & (3119)\\
$(1P,1/2^-)_l$ & 2966 & 3011 & $3015\pm29\pm34$ & &  (3000) \\
$(1P,1/2^-)_h$ & 3055 & 3028 &  &  \\
$(1P,3/2^-)_l$ & 3029 & 2976 &  &  \\
$(1P,3/2^-)_h$ & 3054 & 2993 &  &  & (3055) \\
$1P,5/2^-$ & 3051 & 2947 &  &  & (3066) \\
\end{tabular}
\end{ruledtabular}
\end{table}

\begin{table}[t]
\caption{The $P$-wave $\Omega_c$ baryons denoted by $\Omega_{cJ_\ell}(J^P)$ and $\tilde \Omega_{cJ_\ell}(J^P)$
with $J_\ell$ being the total angular momentum of
the two light quarks \cite{CC,Zhu}. } \label{tab:pwave}
\begin{center}
\begin{tabular}{|c|cccc|} \hline
~~~~~State~~~~~ & SU(3) & ~~$S_\ell$~~ & ~~$L_\ell(L_\rho,L_\lambda)$~~&
~~$J_\ell^{P_\ell}$~~ \\
 \hline
 $\Omega_{c0}({1\over 2}^{-})$ & ${\bf 6}$ & 1 & 1\,(0,1) & $0^-$
 \\
 $\Omega_{c1}({1\over 2}^{-},{3\over 2}^{-})$ & ${\bf 6}$ & 1 & 1\,(0,1) & $1^-$
 \\
 $\Omega_{c2}({3\over 2}^{-},{5\over 2}^{-})$ & ${\bf 6}$ & 1 & 1\,(0,1) & $2^-$
 \\
 $\tilde \Omega_{c1}({1\over 2}^{-},{3\over 2}^{-})$ & ${\bf 6}$ & 0 & 1\,(1,0) & $1^-$
 \\
 \hline
\end{tabular}
\end{center}
\end{table}

\begin{figure}[t]
\begin{center}
\includegraphics[width=100mm]{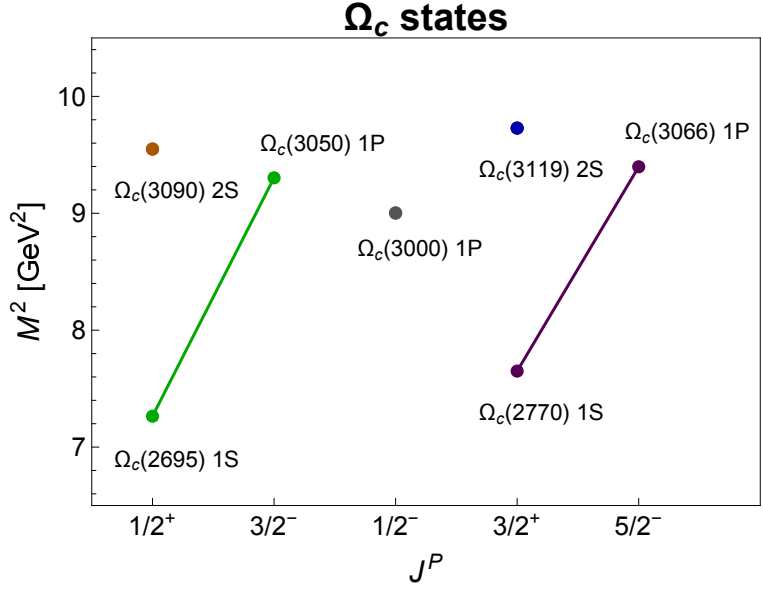}
\caption{Regge trajectories of the $\Omega_c$ states in the $(J^P,M^2)$ plane with natural $(1/2^+,3/2^-)$ and unnatural $(1/2^-,3/2^+,5/2^-)$ parities.}
\label{fig:Omegac}
\end{center}
\end{figure}

\begin{figure}[t]
\begin{tabular}{c}
\includegraphics[width=90mm]{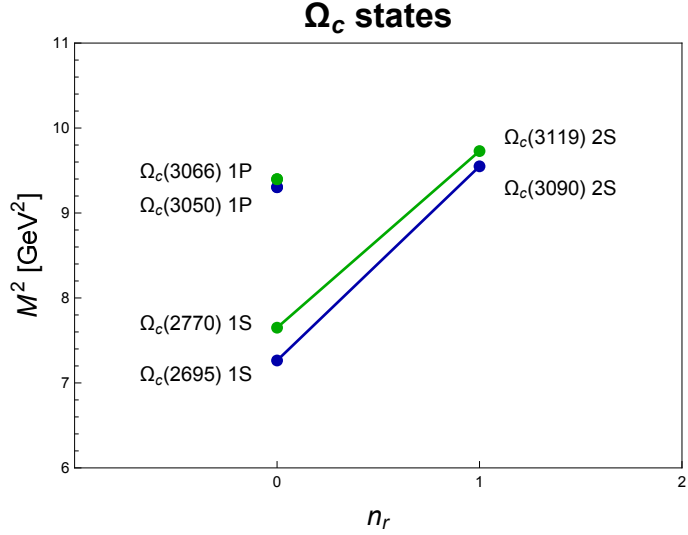}
\end{tabular}
\caption{Regge trajectories of the $\Omega_c$ states in the $(n_r,M^2)$ plane with natural (blue) and unnatural (green) parities.}
\label{fig:nrOmegac}
\end{figure}

Some recent calculations of the $\Omega_c$ spectrum based on the quark model, QCD sum rules, lattice QCD are summarized in Table~\ref{tab:omegac}. (See also Table~6 of \cite{Shah} for a complete compilation of other model predictions.)
Among the five narrow resonances, we can identify the $\frac32^+(2S)$ state with $\Omega_c(3119)$, $(1P,\frac12^-)_l$ with $\Omega_c(3000)$ and $\frac12^+(2S)$ with $\Omega_c(3090)$ from the quark model predictions of \cite{Ebert:2011,Shah}. This is further supported by the lattice QCD calculation for $\Omega_c(3000)$ 
\footnote{A recent lattice calculation with $N_f=2+1+1$ optimal domain-wall fermions \cite{Chiu} yields a mass of $2317\pm15\pm5$ MeV for $D_{s0}^*$ and $2463\pm13\pm9$ MeV for $D'_{s1}(2460)$, in excellent agreement with experiment. It also gives a first lattice result on the mass of the $\frac12^-$ $\Omega_c$ state.}
and QCD sum rules for $\Omega_c(3119)$. Having identified radially excited states of $\Omega_c$ and $\Omega_c^*$, the remaining two resonances $\Omega_c(3050)$ and $\Omega_c(3066)$ should be the orbitally excited states with $J^P={3\over 2}^-$ and ${5\over 2}^-$.
We propose to assign the quantum numbers ${3\over 2}^-$ to $\Omega_c(3050)$ and ${5\over 2}^-$ to $\Omega_c(3066)$.  Such a quantum number assignment is supported by the nearly parallel Regge trajectories of $\Omega_c$ shown in Fig.~\ref{fig:Omegac} and the roughly equal distances between  $\Omega_c(2695)$ and $\Omega_c(3050)$ with natural parities and between  $\Omega_c(2770)$ and $\Omega_c(3066)$ with unnatural parities (see Fig.~\ref{fig:nrOmegac}).

Since many authors \cite{Karliner:2017,Zhao,Wangwei,Wangzg,Padmanath,Chen:Omegac}
claim that the newly observed five $\Omega_c$ resonances can be assigned to the five orbitally excited $1P(1/2^-,3/2^-,5/2^-)$ states, we will go through the details and show that not all the observed $\Omega_c$ baryons can be interpreted as the $P$-wave orbitally excited states.

In the quark model, there are seven first $P$-wave orbitally excited $\Omega_c$ states given in Table \ref{tab:pwave}. Assuming that the spin of the two light quarks $S_\ell$ is 1, a common assumption for the sextet baryons, we are left with five states $\Omega_{c0}({1\over 2}^-)$,  $\Omega_{c1}({1\over 2}^-,{3\over 2}^-)$ and $\Omega_{c2}({3\over 2}^-,{5\over 2}^-)$ in the notation of ${\cal B}_{cJ_\ell}(J^P)$  with $J_\ell$ being the total angular momentum of the two light quarks \cite{CC,Zhu}. The orbital angular momentum of the light
diquark can be decomposed into ${\bf L}_\ell={\bf L}_\rho+{\bf
L}_\lambda$, where ${\bf L}_\rho$ is the orbital angular momentum
between the two light quarks, and ${\bf L}_\lambda$ is the orbital
angular momentum between the diquark and the charmed quark. Denoting the eigenvalues of ${\bf L}_\rho^2$ and
${\bf L}_\lambda^2$
with  $L_\rho$ and $L_\lambda$, respectively, we see that all $\frac12^-(1P)$ $\Omega_c$ states carry $L_\lambda=1$ and $L_\rho=0$.
In the presence of the spin-orbit interaction ${\bf S}_c\cdot{\bf L}$ and the tensor interaction, states with the same $J^P$ but different $J_\ell$ will mix together \cite{Ebert:2011}.
Following \cite{Chen:Sigmac,Chen:Omegac}, we write
\be \label{eq:1/2mixing}
\left(\begin{array}{c} (1P,1/2^-)_l \\ (1P,1/2^-)_h
\end{array}\right)=\left(
\begin{array}{cc} \cos\theta_1 & -\sin\theta_1 \\
\sin\theta_1 & \cos\theta_1 \end{array}\right)
\left(\begin{array}{c} \Omega_{c0}(1/2^-) \\ \Omega_{c1}(1/2^-)
\end{array}\right),
\en
and
\be \label{eq:3/2mixing}
\left(\begin{array}{c} (1P,3/2^-)_h \\ (1P,3/2^-)_l
\end{array}\right)=\left(
\begin{array}{cc} \cos\theta_2 & -\sin\theta_2 \\
\sin\theta_2 & \cos\theta_2 \end{array}\right)
\left(\begin{array}{c} \Omega_{c1}(3/2^-) \\ \Omega_{c2}(3/2^-)
\end{array}\right).
\en
We shall see below that the $(\frac32^-,\frac52^-)$ doublets also exist in $\Sigma_c$ and $\Xi'_c$ sextet states. The mass splitting in the doublet is small and the $3/2^-$ one is slightly heavier than the $5/2^-$ one for $\Sigma_c$ and $\Xi'_c$ sextets.

The strong decays of charmed baryons are most conveniently
described by heavy hadron chiral perturbation theory (HHChPT), into which heavy quark symmetry and chiral symmetry are incorporated \cite{Yan,Wise}. In this approach, the partial widths read
\footnote{Eq.~(\ref{eq:width}) is derived from the Lagrangian terms \cite{Pirjol}
\be
{\cal L} &=& ih_{10} \epsilon_{ijk} \bar \T_i\left( {\cal D}_\mu A_\nu +
{\cal D}_\nu A_\mu\right)_{jl} \X_{kl}^{\mu\nu} +
h_{11}\epsilon_{\mu\nu\sigma\lambda}\mbox{Tr}\left\{ \bar
\S^{\mu}\left( {\cal D}^\nu A_\alpha + {\cal D}_\alpha
A^\nu\right)\X^{\alpha\sigma}
\right\}v^\lambda,  \non
\en
where $\T_i$ and $\S_\mu^{ij}$ are superfields for $S$-wave baryons and $\X^{ij}_{\mu\nu}$ for spin-$\frac52$ and spin-$\frac32$ $J_\ell^{P_\ell}=2^-$ multiplet (see \cite{Pirjol} for details).
}
\begin{align}
\begin{split}
\label{eq:width}
& \Gamma\left(\Omega_{c0}(1/2^-)\to\Xi_c \ov K\right)
 =  {h_{3}^2\over 2\pi f_\pi^2}\,{m_{\Xi_c}\over
 m_{\Omega_{c0}}}E_K^2p_K,
\\
& \Gamma\left(\Omega_{c1}(1/2^-)\to\Xi'_c \ov K\right)
 =  {h_{4}^2\over 4\pi f_\pi^2}\,{m_{\Xi'_c}\over
 m_{\Omega_{c1}}}E_K^2p_K,
\\
& \Gamma\left(\Omega_{c1}(3/2^-)\to\Xi'_c \ov K\right)
 =  {h_{9}^2\over 9\pi f_\pi^2}\,{m_{\Xi'_c}\over
 m_{\Omega_{c1}}}p_K^5,
\\
& \Gamma\left(\Omega_{c2}(3/2^-,5/2^-)\to\Xi_c \ov K\right)
 =  {4h_{10}^2\over 15\pi f_\pi^2}\,{m_{\Xi_c}\over
 m_{\Omega_{c2}}}p_K^5,
\\
& \Gamma\left(\Omega_{c2}(3/2^-)\to\Xi'_c \ov K\right) =
 {h_{11}^2\over 10\pi f_\pi^2}\,{m_{\Xi'_c}\over
 m_{\Omega_{c2}}}p_K^5,
\\
& \Gamma\left(\Omega_{c2}(5/2^-)\to\Xi'_c \ov K\right) =
 {2h_{11}^2\over 45\pi f_\pi^2}\,{m_{\Xi'_c}\over
 m_{\Omega_{c2}}}p_K^5,
\end{split}
\end{align}
where $p_K$ is the center-of-mass (c.m.) momentum of the kaon and $f_\pi=132$~MeV.  In the above equations, $h_{3,4}$ are the couplings responsible for the $s$-wave transition between $S$- and $P$-wave baryons and $h_{9,10,11}$ are the couplings for the $d$-wave transition between $S$- and $P$-wave baryons. Using the quark model relation $|h_3|=\sqrt{3}|h_2|$ from \cite{Pirjol} and the coupling $h_2$ extracted from $\Lambda_c(2595)^+\to \Lambda_c^+\pi^+\pi^-$, it is found that $\Gamma(\Omega_{c0}\to\Xi_c \ov K)\approx 410$ MeV for $h_2=0.437$ \cite{CC} and 852 MeV for $h_2=0.63$ \cite{CC:2015}.~
\footnote{The coupling $h_2$ was used to be of order 0.42\,. It became large, of order 0.60, after a more sophisticated treatment of the mass lineshape of $\Lambda_c(2595)^+\to\Lambda_c^+\pi^+\pi^-$ by the CDF~\cite{CDF:2595}. However, this latest value of $h_2$ will lead to the predictions of $\Gamma(\Xi_c^+(2790))$ and $\Gamma(\Xi_c^0(2790))$ too large by a factor of 2 compared to the recent measurements by Belle~\cite{Belle:Xic}. Therefore, we should use $h_2=0.437^{+0.114}_{-0.102}$ \cite{CC} in the ensuing discussions.
}
Hence, $\Omega_c(3000)$ cannot be a pure $\Omega_{c0}(\frac12^-)$ state due to a very broad width expected for the $s$-wave transition. Nevertheless, it can be identified with $\Omega_{c1}(\frac12^-)$ since its decay into $\Xi_c K$ is prohibited in the heavy quark limit but could be allowed when heavy quark symmetry is broken. This means that the mixing angle $\theta_1$ in Eq. (\ref{eq:1/2mixing}) must be close to $90^\circ$ if $\Omega_c(3000)$ is to be identified with the $(1P,1/2^-)_l$ state.
From the data $\Gamma(\Omega_c(3000))=4.5\pm0.7$ MeV \cite{LHCb:Omegac}, we find that $\theta_1\approx 96^\circ$ or $84^\circ$ where we have neglected the contributions from $\Omega_{c1}(\frac12^-)\to\Xi_c\ov K$.
The other state $(1P,1/2^-)_h$ will be too broad to be observed. For example, if we identify $\Omega_c(3090)$ with $(1P,1/2^-)_h$, we will obtain $\Gamma(\Omega_c(3090)\to \Xi_c\ov K+\Xi'_c\ov K)=\sin^2\theta_1(1006\,{\rm MeV})+\cos^2\theta_1(173\,{\rm MeV})=997$ MeV for $\theta_1=96^\circ$, where use of $|h_4|=2|h_2|$ \cite{Pirjol} has been made.
Hence, we conclude that only one of the  $(1P,1/2^-)$ states can be identified with the observed narrow $\Omega_c$ baryon. We see that $\Omega_c(3000)$ is narrow because it is primarily a $\Omega_{c1}(\frac12^-)$ state with a very small component of $\Omega_{c0}(\frac12^-)$.

We next turn to the widths of $\Omega_c(3050)$ and $\Omega_c(3066)$. It is clear from Eq. (\ref{eq:width}) that their widths are governed by the coupling $h_{10}$
which can be determined from the measured widths of $\Sigma_c(2800)^{++,+,0}$ to be \cite{CC:2015}
 \begin{eqnarray} \label{eq:h10}
|h_{10}|=(0.85^{+0.11}_{-0.08})\times 10^{-3}\,{\rm MeV}^{-1}\,.
 \end{eqnarray}
We then obtain $\Gamma(\Omega_c(3050))=\sin^2\theta_2(8.6^{+2.2}_{-1.6})$~MeV and $\Gamma(\Omega_c(3066))=(13.3^{+3.4}_{-2.5})$~MeV where we have neglected the contribution from $\Omega_{c1}(3/2^-)$ as it does not decay into $\Xi_cK$ in the heavy quark limit.  The experimental width of $(0.8\pm0.2\pm0.1)<1.2$ MeV for $\Omega_c(3050)$ \cite{LHCb:Omegac} is well accommodated for $\theta_2\approx 160^\circ$, but
our prediction for $\Omega_c(3066)$ is too large by a factor of 4 compared to the data $3.5\pm0.4\pm0.2$ MeV   \cite{LHCb:Omegac}. It is not clear to us what is the underlying reason for this discrepancy. For example, lowering the estimate of background events in the data may bring the observed widths closer to our calculations.

There are two recent papers claiming reasonable model results for the $\Omega_c$ widths: \cite{Zhao} and \cite{Chen:Omegac}. Using the decay formula proposed by Eichten, Hill and Quigg and the $^3P_0$ model in conjunction with the simple harmonic oscillator wave functions for the transition form factors, Chen and Liu \cite{Chen:Omegac} calculated partial and total widths for the $1P$ and $2S$ $\Omega_c$ states. They obtained $\Gamma(\Omega_{c0})=35$ MeV (see Fig.~1 of \cite{Chen:Omegac}), which was smaller than our model-independent result by one order of magnitude. For comparison, we notice that a very broad width of 1400 MeV for $\Omega_{c0}$ is predicted in \cite{Zhao:Omegac}, while  the QCD sum rule result of 420 MeV \cite{Chen:Pwave} is very close to ours. As noticed in passing, if the width of $\Omega_{c0}$ is indeed of order 400 MeV, not both $(1P,1/2^-)_l$ and $(1P,1/2^-)_h$ can be identified with the observed narrow $\Omega_c$ states.  Wang {\it et al.} computed the strong and radiative decays of $\Omega_c$ states using the chiral quark model \cite{Zhao} and obtained narrow widths for all $^{2S+1}L_{J^P}$ states for $L=1$, $J^P=1/2^-,3/2^-$ and $5/2^-$. In this work, the authors did not consider the mixing effects of the states with the same $J$ but different $J_\ell$ or $S$. We suspect that at least some widths calculated in \cite{Zhao} and \cite{Chen:Omegac} are underestimated.

\subsection{$\Lambda_c$ states}

\begin{figure}[t]
\begin{center}
\includegraphics[width=100mm]{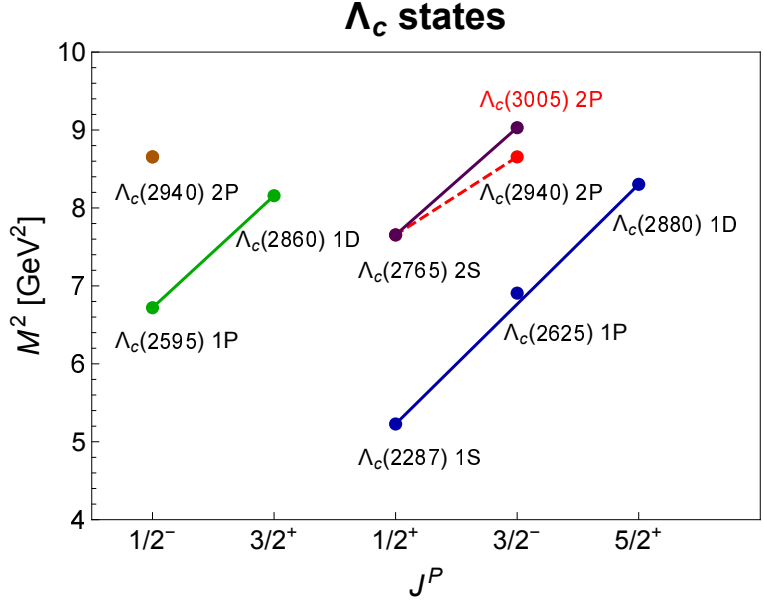}
\caption{Regge trajectories of the $\Lambda_c$ states in the $(J^P,M^2)$ plane with natural $(1/2^+,3/2^-,5/2^+)$ and unnatural $(1/2^-,3/2^+)$ parities.  The yet detected state is labeled in red.  The red dashed line shows a discordant identification of $\Lambda_c(2940)$ as a $3/2^-(2P)$ state.}
\label{fig:Lambdac}
\end{center}
\end{figure}

$\Lambda_c(2765)^+$ is a broad state first seen in the $\Lambda_c^+\pi^+\pi^-$ decay by CLEO~\cite{CLEO:Lamc2880}. However, it is still not known whether it is
$\Lambda_c^+$ or $\Sigma_c^+$ and whether the large width might be due
to overlapping states.  In the quark-diquark model, it has also been proposed to be either the first radial ($2S$)
excitation of the $\Lambda_c$ with $J^P=\frac12^-$ containing the light scalar diquark  or the first orbital
excitation ($1P$) of the $\Sigma_c$ with $J^P=\frac32^-$ containing the light axial-vector diquark \cite{Ebert:2007}. In this work we shall consider the former case.

The state $\Lambda_c(2880)^+$, first observed by CLEO~\cite{CLEO:Lamc2880} in the $\Lambda_c^+\pi^+\pi^-$ decay, was also seen by
BaBar in the $D^0p$ spectrum~\cite{BaBar:Lamc2940}.  Belle studied the
experimental constraint on the $J^P$ quantum numbers of
$\Lambda_c(2880)^+$~\cite{Belle:Lamc2880} and found that $J^P=\frac52^+$ was favored by the angular analysis of $\Lambda_c(2880)^+\to\Sigma_c^{0,++}\pi^\pm$ decays. The mass, width and quantum numbers of $\Lambda_c(2880)$ were recently confirmed by LHCb~\cite{LHCb:Lambdac2880}. The $\frac12^+(1S)$ $\Lambda_c$, $\frac32^-(1P)$ $\Lambda_c(2625)$ and $\frac52^+(1D)$ $\Lambda_c(2880)$ states form a Regge trajectory. The new resonance $\Lambda_c(2860)^+$ observed by LHCb, as manifested in the near-threshold enhancement in the $D^0p$ amplitude through an amplitude analysis of the $\Lambda_b^0\to D^0p\pi^-$ decay, has $J^P=\frac 32^+$ with mass and width shown in Table~\ref{tab:spectrum}~\cite{LHCb:Lambdac2880}.
It forms another Regge trajectory with $\frac12^-(1P)$ $\Lambda_c(2595)$.
It is worth mentioning that the existence of this new state $\Lambda_c(2860)^+$ was noticed  before the LHCb experiment~\cite{Chen:Sigmac,Lu:2016ctt,Chen:Dwave}.
We see from Fig.~\ref{fig:Lambdac} that both trajectories are parallel to each other nicely.

The highest state $\Lambda_c(2940)^+$ was first discovered by BaBar in
the $D^0p$ decay mode~\cite{BaBar:Lamc2940} and  confirmed by
Belle in the $\Sigma_c^0\pi^+,\Sigma_c^{++}\pi^-$ decays, which
subsequently decayed into $\Lambda_c^+\pi^+\pi^-$~\cite{Belle:Lamc2880}. Its spin-parity assignment is quite diverse (see \cite{Cheng:2015} for a review). The constraints on its spin and parity were recently studied by LHCb~\cite{LHCb:Lambdac2880}. The most likely assignment was found to be $J^P=\frac32^-$ with
\begin{align}
\begin{split}
m(\Lambda_c(2940)) &= 2944.8^{+3.5}_{-2.5}\pm0.4^{+0.1}_{-4.6}~{\rm MeV},
\\
\Gamma(\Lambda_c(2940)) &= 27.7^{+8.2}_{-6.0}\pm0.9^{+~5.2}_{-10.4}~{\rm MeV},
\end{split}
\end{align}
to be compared with $m=2939.3^{+1.4}_{-1.5}$ MeV and $\Gamma=17^{+8}_{-6}$ MeV quoted in PDG \cite{PDG}.  We have averaged them in Table \ref{tab:spectrum}.
If we draw a Regge trajectory connecting $\Lambda_c(2940)$ and $\Lambda_c(2765)$ with $\frac12^+(2S)$, we see that this Regge line is not parallel to the other two Regge trajectories. If we use the quark-diquark model prediction of $\Lambda_c(3005)$ for the $\frac32^-(2P)$ state \cite{Ebert:2011}, the trajectories satisfy the parallelism nicely. Hence, we suggest that the quantum numbers of $\Lambda_c(2940)^+$ are most likely $\frac12^-(2P)$.
Indeed, LHCb has cautiously stated that ``The most likely spin-parity assignment for $\Lambda_c(2940)$ is $J^P=\frac32^-$ but the other solutions with spin 1/2 to 7/2 cannot be excluded.'' In order to clarify this issue, it is thus important to search for the $\Lambda_c^+$ state with a mass of order 3005 MeV and verify its quantum numbers as $\frac32^-(2P)$.

\subsection{$\Xi_c$ states}

\begin{figure}[t]
\begin{center}
\vspace{10pt}
\includegraphics[width=100mm]{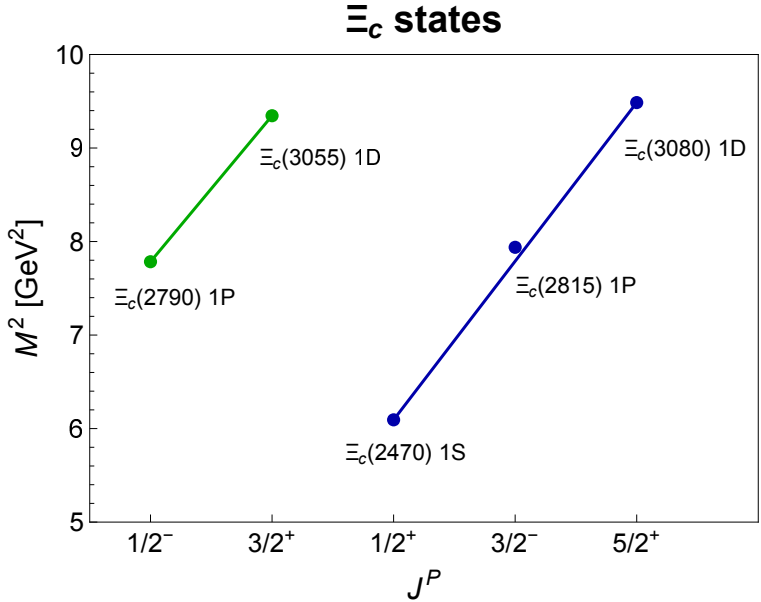}
\caption{Regge trajectories of the $\Xi_c$ states in the $(J^P,M^2)$ plane  with natural $(1/2^+,3/2^-,5/2^+)$ and unnatural $(1/2^-,3/2^+)$ parities.}
\label{fig:Xic}
\end{center}
\end{figure}

Another example showing the usefulness of the Regge phenomenology in the $J^P$ assignment of charmed baryons is the $\Xi_c$ states.
The Regge analysis suggests $3/2^+(1D)$ for $\Xi_c(3055)$ and $5/2^+(1D)$ for $\Xi_c(3080)$~\cite{Ebert:2011} (see also discussions in~\cite{Chen:Lambdac2880}). The $\Xi_c(2470)$, $\Xi_c(2815)$ and $\Xi_c(3080)$ states form a $\frac12^+$ Regge trajectory, while  $\Xi_c(2790)$ and $\Xi_c(3055)$ form a $\frac12^-$ one (see Fig.~\ref{fig:Xic}). They are parallel to each other nicely.
Recently, the discovery of the neutral $\Xi_c(3055)^0$, observed by its decay into the final-state $\Lambda D^0$, and the first observation and evidence of the decays of $\Xi_c(3055)^+$ and $\Xi_c(3080)^+$ into $\Lambda D^+$ were presented by Belle~\cite{Belle:Kato}.

\subsection{$\Xi'_c$ states}

\begin{figure}[t]
\begin{center}
\includegraphics[width=100mm]{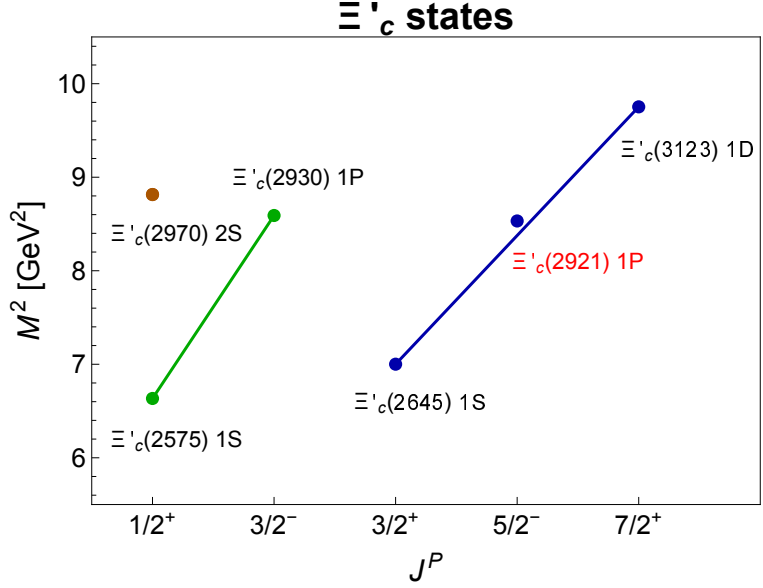}
\caption{Regge trajectories of the $\Xi'_c$ states in the $(J^P,M^2)$ plane  with natural $(1/2^+,3/2^-)$ and unnatural $(3/2^+,5/2^-,7/2^+)$ parities.  The yet detected state is labeled in red.}
\label{fig:Xipc}
\end{center}
\end{figure}

A state $\Xi_c(2930)^0$, which is omitted from the PDG summary table, has been seen only by BaBar in the $\Lambda_c^+K^-$ mass projection of $B^-\to\Lambda_c^+\bar\Lambda_c^-K^-$ \cite{BaBar:Xic2930}. According to the quark-diquark model of~\cite{Ebert:2011} (see also~\cite{Chen:Sigmac}), its $J^P$ quantum numbers could be $\frac32^-$ or $\frac52^-$.  Quark model calculations suggest that $\frac32^-(1P)$ is slightly heavier than $\frac52^-(1P)$ (see Table~3 of~\cite{Chen:Sigmac}). The $\Xi'_c(2645)$  state with $\frac32^+(1S)$ and $\Xi'_c(3123)$ with $\frac72^+(1D)$ form a Regge trajectory. It is clear from Fig.~\ref{fig:Xipc} that the unknown $\frac52^-$ state has a mass of order 2890 MeV. We shall designate  this state to $\Xi'_c(2921)$ which carries the correct spin-parity quantum numbers and its mass is not far from 2890~MeV~\cite{Ebert:2007}.  Hence, we should assign $\frac32^-$ to $\Xi'_c(2930)$. Now $\Xi'_c(2930)$ and $\Xi'_c(2921)$ form a $P$-wave doublet denoted by $\Xi'_{c2}(\frac32^-,\frac52^-)$. Just as the $\Omega_{c2}$ doublet, the partial widths of $\Xi'_{c2}(3/2^-)$ read
\begin{align}
\begin{split}
&
 \Gamma\left(\Xi'_{c2}(3/2^-)\to\Lambda_c \ov K\right)
 =  {4h_{10}^2\over 15\pi f_\pi^2}\,{m_{\Lambda_c}\over
 m_{\Xi'_{2c}}}p_K^5,
~~
 \Gamma\left(\Xi'_{c2}(3/2^-)\to\Xi_c \pi\right)
 =  {4h_{10}^2\over 15\pi f_\pi^2}\,{m_{\Xi_c}\over
 m_{\Xi'_{2c}}}p_\pi^5,
\\
&
 \Gamma\left(\Xi'_{c2}(3/2^-)\to\Sigma_c \ov K\right) =
 {h_{11}^2\over 10\pi f_\pi^2}\,{m_{\Sigma_c}\over
 m_{\Xi'_{c2}}}p_K^5,
~~
\Gamma\left(\Xi'_{c2}(3/2^-)\to\Xi'_c \pi\right) =
 {h_{11}^2\over 10\pi f_\pi^2}\,{m_{\Xi'_c}\over
 m_{\Xi'_{c2}}}p_\pi^5.
\end{split}
\end{align}
If the state $\Xi'_{c2}(3/2^-)$ is identified with $\Xi'_c(2930)$, its decay into $\Sigma_c\ov K$ will be kinematically prohibited. Although $\Xi'_c(2930)$ has been observed only in the $\Lambda_cK$ decay mode, we need to sum over the $\Lambda_c\ov K, \Xi_c\pi,\Xi'_c\pi$ channels in order to estimate its total width. Using the quark model relation $h_{11}^2=2h^2_{10}$~\cite{Pirjol} and Eq.~(\ref{eq:h10}), we obtain
\be
\Gamma(\Xi'_c(2930)^0)=77^{+20}_{-14}\,{\rm MeV},
\en
which deviates from the measurement of $36\pm13$ MeV  \cite{BaBar:Xic2930} by $2.1\sigma$.  One possibility for the discrepancy is ascribed to the SU(3) breaking in the quark model relation $h_{11}^2=2h^2_{10}$.
In view of theoretical difficulties in estimating decay widths, we regard the above HHChPT result as a good support for the $\frac32^-(1P)$ assignment to $\Xi'_c(2930)$.

\subsection{$\Sigma_c$ states}

\begin{figure}[t]
\begin{center}
\includegraphics[width=100mm]{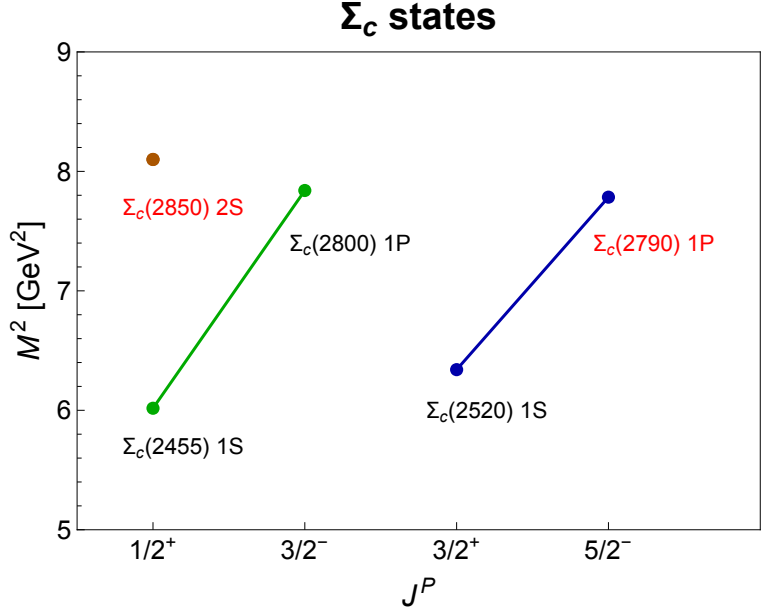}
\caption{Regge trajectories of the $\Sigma_c$ states in the $(J^P,M^2)$ plane  with natural $(1/2^+,3/2^-)$ and unnatural $(3/2^+,5/2^-)$ parities.  The yet detected states are labeled in red.}
\label{fig:Sigmac}
\end{center}
\end{figure}

The highest isotriplet charmed baryons, $\Sigma_c(2800)^{++,+,0}$,
decaying to $\Lambda_c^+\pi$, were first measured by Belle
\cite{Belle:Sigc2800} with widths of order 70~MeV.
We have advocated in \cite{CC} that they are $\Sigma_{c2}(\frac32^-)$ states. Their quantum numbers are sometimes assigned to be $\frac12^-$ in the literature. Here we repeat our argument again.
The possible quark states are $\Sigma_{c0}({1\over 2}^-)$,  $\Sigma_{c1}({1\over 2}^-,{3\over 2}^-)$, $\tilde\Sigma_{c1}({1\over 2}^-,{3\over 2}^-)$, and $\Sigma_{c2}({3\over 2}^-,{5\over 2}^-)$ in the notation of ${\cal B}_{cJ_\ell}(J^P)$~\cite{CC,Zhu}, or $[{\bf 6}_F,0,1,\lambda]$,  $[{\bf 6}_F,1,1,\lambda]$,  $[{\bf 6}_F,1,0,\rho]$ and $[{\bf 6}_F,2,1,\lambda]$
in terms of the notation
$[{\bf 6}_F,J_\ell,S_\ell, \rho/\lambda]$.
The states $\Sigma_{c1}$ and $\tilde\Sigma_{c1}$ are ruled out
because their decays to $\Lambda_c^+\pi$ are prohibited in the heavy quark limit, recalling that only the $\Sigma_c(2800)\to\Lambda_c\pi$ decay mode has been observed.  Now the
$\Sigma_{c2}({3\over 2}^-,{5\over 2}^-)$ baryons decay primarily into the
$\Lambda_c\pi$ system in a $d$-wave, whereas $\Sigma_{c0}({1\over 2}^-)$  decays into $\Lambda_c\pi$ in an $s$-wave. In the framework of HHChPT, we have \cite{Pirjol}
\be
\Gamma(\Sigma_{c0}(1/2^-)\to \Lambda_c\pi)={h_3^2\over 2\pi
 f_\pi^2}\,{m_{\Lambda_c}\over m_{\Sigma_{c0}}}E_\pi^2p_\pi,
\en
where $h_3$ is one of the couplings responsible for the $s$-wave transition between $S$- and $P$-wave baryons, and $p_\pi$ is the c.m.~momentum of the pion. Using the quark model relation $|h_3|=\sqrt{3}|h_2|$ from \cite{Pirjol} and the coupling $h_2$ extracted from $\Lambda_c(2595)^+\to \Lambda_c^+\pi^+\pi^-$, it is found that $\Gamma(\Sigma_{c0}^{++}\to\Lambda_c^+\pi^+)\approx 425$ MeV for $h_2=0.437$ \cite{CC} and 885 MeV for $h_2=0.63$ \cite{CC:2015}.
In either case, the predicted width is too large by one order of magnitude compared to the measured one of order 75 MeV. Hence, this very broad $\Sigma_{c0}$ cannot be identified with $\Sigma_c(2800)$.
Therefore, $\Sigma_c(2800)^{++,+,0}$ are likely to
be either $\Sigma_{c2}({3\over 2}^-)$ or $\Sigma_{c2}({5\over 2}^-)$ or a mixture of the two. In the quark-diquark model \cite{Ebert:2011}, both of them have very close masses compatible with experiment.  Given the fact that for light strange baryons, the first orbital excitation of the light $\Sigma$ has the quantum numbers $J^P=\frac32^-$, we thus advocate a $\Sigma_{c2}(\frac32^-)$ state for $\Sigma_c(2800)$. The $\frac52^-$ $\Sigma_c(2790)$ state has a mass in the vicinity of 2790~MeV \cite{Chen:Sigmac,Ebert:2011}.

Using QCD sum rules, the authors of \cite{Chen:Pwave} obtained the widths of 200~MeV, 7.9~MeV and 300~MeV respectively for the $\Sigma_{c0}(\frac12^-)\to \Lambda_c\pi$,
$\Sigma_{c1}(\frac12^-)\to \Sigma_c\pi$ and $\tilde\Sigma_{c1}(\frac12^-)\to \Sigma_c\pi$ decays, and proposed that $\Sigma_c(2800)$ might be a $\frac12^-$ state belonging to $\Sigma_{c0}$ or as a $\frac12^-$ state containing both $\Sigma_{c0}$ and $\Sigma_{c1}$.

Among the sextet states, both $\Omega_c$ and $\Xi'_c$ have $\frac12^+(2S)$ states: $\Omega_c(3090)$ and $\Xi'_c(2970)$. In the $\Sigma_c$ sector, we also have a possible $\frac12^+(2S)$ candidate. BaBar observed an excited $\Sigma_c^0$ state (denoted as $\Sigma_c^0(2850)$ in \cite{Chen:Sigmac}) in the decay $B^-\to\Sigma_c(2850)^0\bar p\to\Lambda_c^+\pi^-\bar p$ with a mass of $2846\pm8\pm10$~MeV and a width of $86^{+33}_{-22}$~MeV~\cite{BaBar:Sigmac2846}. We shall follow \cite{Chen:Sigmac} to designate this new state with $\frac12^+(2S)$.
Regge trajectories for the $\Sigma_c$ states are plotted in Fig.~\ref{fig:Sigmac}.

\subsection{Antitriplet and sextet states}

\begin{table}[t]
\caption{Antitriplet and sextet states of charmed baryons.
Mass differences $\Delta m_{\Xi_c\Lambda_c}\equiv m_{\Xi_c}-m_{\Lambda_c}$, $\Delta m_{\Xi'_c\Sigma_c}\equiv m_{\Xi'_c}-m_{\Sigma_c}$, $\Delta m_{\Omega_c\Xi'_c}\equiv m_{\Omega_c}-m_{\Xi'_c}$ are all in units of MeV. The yet detected states are labeled in red.} \label{tab:3and6}
\begin{center}
\begin{tabular}{|c| ccc |} \hline\hline
  & $J^P(nL)$ & States & Mass differences  \\
 \hline
 ~~${\bf \bar 3}$~~ & ~~${1\over 2}^+(1S)$~~ &  $\Lambda_c(2287)^+$, $\Xi_c(2470)^+,\Xi_c(2470)^0$ & ~~$\Delta m_{\Xi_c\Lambda_c}=183$ ~~  \\
 & ~~${1\over 2}^-(1P)$~~ &  $\Lambda_c(2595)^+$, $\Xi_c(2790)^+,\Xi_c(2790)^0$ & $\Delta m_{\Xi_c\Lambda_c}=198$  \\
 & ~~${3\over 2}^-(1P)$~~ &  $\Lambda_c(2625)^+$, $\Xi_c(2815)^+,\Xi_c(2815)^0$ & $\Delta m_{\Xi_c\Lambda_c}=190$  \\
 & ~~${3\over 2}^+(1D)$~~ &  $\Lambda_c(2860)^+$, $\Xi_c(3055)^+,\Xi_c(3055)^0$ & $\Delta m_{\Xi_c\Lambda_c}=201$  \\
 & ~~${5\over 2}^+(1D)$~~ &  $\Lambda_c(2880)^+$, $\Xi_c(3080)^+,\Xi_c(3080)^0$ & $\Delta m_{\Xi_c\Lambda_c}=196$  \\
 \hline
 ~~${\bf 6}$~~ & ~~${1\over 2}^+(1S)$~~ &  $\Omega_c(2695)^0$, $\Xi'_c(2575)^{+,0},\Sigma_c(2455)^{++,+,0}$ & ~~~~$\Delta  m_{\Omega_c\Xi'_c}=119$, $\Delta m_{\Xi'_c\Sigma_c}=124$~~  \\
 & ~~~${3\over 2}^+(1S)$~~~ &  $\Omega_c(2770)^0$, $\Xi'_c(2645)^{+,0},\Sigma_c(2520)^{++,+,0}$ & ~~~~$\Delta m_{\Omega_c\Xi'_c}=120$, $\Delta m_{\Xi'_c\Sigma_c}=128$~~ \\
 & ~~${1\over 2}^+(2S)$~~ &  $\Omega_c(3090)^0$, $\Xi'_c(2970)^{+,0}, \color{red}{\Sigma_c(2850)}^{++,+,0}$ & ~~~~$\Delta m_{\Omega_c\Xi'_c}=120$, $\Delta m_{\Xi'_c\Sigma_c}=120$~~  \\
 & ~~${3\over 2}^-(1P)$~~ &  $\Omega_c(3050)^0$, $\Xi'_c(2930)^{+,0},\Sigma_c(2800)^{++,+,0}$ & ~~~~$\Delta m_{\Omega_c\Xi'_c}=120$, $\Delta m_{\Xi'_c\Sigma_c}=130$~~  \\
 & ~~${5\over 2}^-(1P)$~~ &  $\Omega_c(3066)^0$, $\color{red}{\Xi'_c(2921)}^{+,0}$, $\color{red}{\Sigma_c(2790)}^{++,+,0}$ & ~~~~$\Delta m_{\Omega_c\Xi'_c}=145$, $\Delta m_{\Xi'_c\Sigma_c}=131$~~  \\
 \hline\hline
\end{tabular}
\end{center}
\end{table}

Many observed charmed baryons form antitriplet and sextet states. They
are classified according to the quantum numbers $J^P(nL)$
in Table~\ref{tab:3and6}.  The mass difference $\Delta m_{\Xi_c\Lambda_c}\equiv m_{\Xi_c}-m_{\Lambda_c}$ in the antitriplet states clearly lies between about 180 and 200~MeV. This means that the quantum numbers of the listed  ${\bf \bar 3}$ states  are now established.  Also shown in Table \ref{tab:3and6} are five different sets of sextet states associated with the $\Omega_c,\Xi'_c$ and $\Sigma_c$ baryons. The states labeled in red are yet to be measured and have been discussed in previous subsections.
The mass splittings  $\Delta m_{\Omega_c\Xi'_c}\equiv m_{\Omega_c}-m_{\Xi'_c}$ between $\Omega_c$ and $\Xi'_c$ and $\Delta m_{\Xi'_c\Sigma_c}\equiv m_{\Xi'_c}-m_{\Sigma_c}$ between $\Xi'_c$ and $\Sigma_c$ ought to be about the same. Numerically, we find that $\Delta m_{\Omega_c\Xi'_c}$  and $\Delta m_{\Xi'_c\Sigma_c}$ are indeed close to each other, between about 120 and 130~MeV. This lends further a strong support for the quantum number assignment to the sextet states in this work.

It is clear from Table~\ref{tab:3and6} that various doublets are observed. In the antitriplet sector, $(\Lambda_c(2595),\Lambda_c(2625))$ and $(\Xi_c(2790),\Xi_c(2815))$ belong to the $P$-wave doublets $(\frac12^-,\frac32^-)$ while $(\Lambda_c(2860),\Lambda_c(2880))$ and $(\Xi_c(3055),\Xi_c(3080))$ form the $D$-wave doublets $(\frac32^+,\frac52^+)$.  In the sextet sector, ($\Omega_c(2695),\Omega_c(2770))$, $(\Sigma_c(2455),\Sigma_c(2520))$ and $(\Xi'_c(2575),\Xi'_c(2645))$ belong to the $S$-wave doublets $(\frac12^+,\frac32^+)$ while ($\Omega_c(3050),\Omega_c(3066))$, $(\Sigma_c(2800),\Sigma_c(2790))$ and $(\Xi'_c(2930),\Xi'_c(2921))$ form the $P$-wave doublets $(\frac32^-,\frac52^-)$.

\subsection{Regge trajectories}

Various Regge trajectories in the $(J^P,M^2)$ plane for $\Omega_c,\Lambda_c,\Xi_c,\Xi'_c$ and $\Sigma_c$ states are depicted in Figs.~\ref{fig:Omegac} to \ref{fig:Sigmac}. In the phenomenology of Regge trajectories, the Regge slopes are usually assumed to be the same for all the baryon multiplets. This ansatz leads to the parallelism among trajectories with natural or unnatural parities, and the parallelism
between natural and unnatural parities.
Empirically, this is nicely supported by the Regge trajectories of
the antitriplet $\Lambda_c$ and $\Xi_c$ states. We see that their Regge trajectories for the orbital excitations of $\frac12^-$ and $\frac32^-$ are parallel to each other, as shown in Figs.~\ref{fig:Lambdac} and \ref{fig:Xic}. Based on this nice property of parallelism, we have shown that the quantum numbers of $\Lambda_c(2940)^+$ are most likely $\frac12^-(2P)$ rather than $\frac32^-$ found by LHCb \cite{LHCb:Lambdac2880}.

As for the sextet $\Omega_c$, $\Xi'_c$ and $\Sigma_c$ states, the slope of the Regge trajectory for the orbital excitation of $\frac12^+$ is slightly larger than that of the $\frac32^+$ one for reasons not clear to us. The $\frac32^-$ and $\frac52^-$ states $(\Omega_c(3050),\Omega_c(3066))$, $(\Xi'_c(2930),\Xi'_c(2921))$
and $(\Sigma_c(2800),\Sigma_c(2790))$ form $P$-wave doublets described by $[{\bf 6}_F,2,1,\lambda]$ or $\Omega_{c2}(\frac32^-,\frac52^-),\Xi'_{c2}(\frac32^-,\frac52^-),\Sigma_{c2}(\frac32^-,\frac52^-)$, respectively. The mass splittings in the doublets are small and the $\frac32^-$ states are slightly heavier than the $\frac52^-$ ones.

\begin{figure}[t]
\begin{tabular}{c}
\includegraphics[width=90mm]{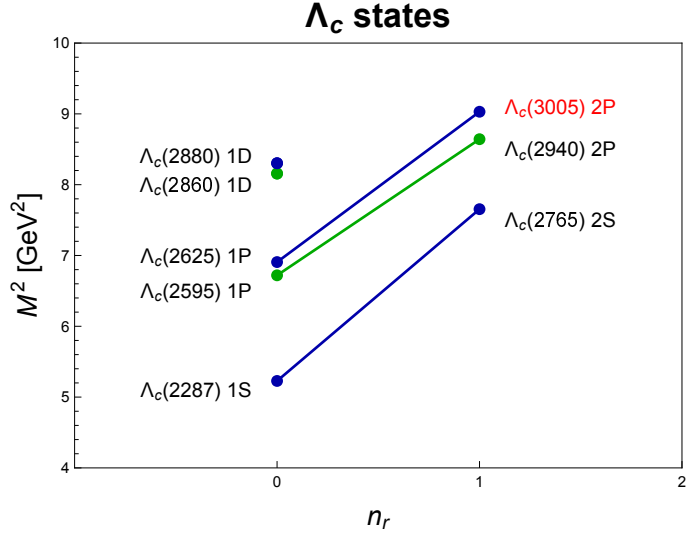}
\end{tabular}
\caption{Regge trajectories of the $\Lambda_c$ states in the $(n_r,M^2)$ plane with natural (blue) and unnatural (green) parities.}
\label{fig:nrLambdac}
\end{figure}

For completeness, we also show the Regge trajectories in the $(n_r,M^2)$ plane for $\Omega_c$ and $\Lambda_c$ in Figs.~\ref{fig:nrOmegac} and \ref{fig:nrLambdac}, respectively. The parallelism and nearly equidistance of the Regge trajectories of $\Lambda_c$ states with natural parities $(1/2^+,3/2^-,5/2^+)$ are obviously seen in Fig.~\ref{fig:nrLambdac}.

\section{Conclusions}

Based mainly on the heavy quark-light diquark model and the Regge trajectories in conjunction with other model calculations, we have studied the spin-parity quantum numbers of charmed baryons. Our main results are as follows:

\begin{itemize}
\item
Among the five newly observed $\Omega_c$ states, we have identified $\Omega_c(3090)$ and $\Omega_c(3119)$ with the radially excited  $\frac12^+(2S)$ and $\frac32^+(2S)$ states, respectively, and $\Omega_c(3000)$ with $\frac12^-(1P)$ and $S=\frac32$. The two states $\Omega_c(3050)$ and $\Omega_c(3066)$ form a $P$-wave $(\frac32^-,\frac52^-)$ doublet.

\item
Since the width of $\Omega_{c0}(\frac12^-)$ is estimated to be of order 410 MeV using heavy hadron chiral perturbation theory, not all observed narrow $\Omega_c$ baryons can be identified with $1P$ states. The mixing angles $\theta_1$ and $\theta_2$ defined in Eqs.~(\ref{eq:1/2mixing}) and (\ref{eq:3/2mixing}) are constrained to be around $96^\circ$ and $160^\circ$, respectively. 

\item
In the sextet sector, $(\Sigma_c(2800),\Sigma_c(2790))$ and $(\Xi'_c(2930),\Xi'_c(2921))$ also belong to the $P$-wave $(\frac32^-,\frac52^-)$ doublet. Using the measured width of $\Sigma_c(2800)$ as an input, the widths of $\Omega_c(3050)$, $\Omega_c(3066)$ and $\Xi'_c(2930)$ are calculable within the framework of heavy hadron chiral perturbation theory. The predicted width of $\Xi'_c(2930)$ deviates from experiment by $2.1\sigma$. While $\Omega_c(3066)$ is broader than $\Omega_c(3050)$, it is narrower than $\Sigma_c(2880)$ and $\Xi'_c(2930)$ by one order of magnitude due to the smaller c.m.~momentum $p_K$ appearing in the $D$-wave suppression factor proportional to $p_K^5$.

\item
For the $\Lambda_c$ and $\Xi_c$ antitriplet states, their Regge trajectories for the orbital excitations of $\frac12^-$ and $\frac32^-$ are parallel to each other. Based on this nice property of parallelism, we see that although the newly detected $\Lambda_c(2860)^+$ fits nicely to the Regge trajectory, the highest state $\Lambda_c(2940)^+$ does not fit if its quantum numbers are $\frac32^-$ as preferred by LHCb.  We suggest that $\Lambda_c(2940)^+$ is most likely the $\frac12^-(2P)$ state. Experimentally, it is thus important to search for the $\Lambda_c$ baryon with a mass of order 3005 MeV and verify its quantum numbers as $\frac32^-(2P)$.

\item
The charmed baryon $\Sigma_c(2800)$ cannot be a $\frac12^-$ state. Otherwise, its width will be over 400~MeV, too large compared to the measured one.

\item
In the study of Regge trajectories of $\Xi'_c$ states, we find a missing state. It should have quantum numbers $\frac52^-$ with a mass around 2920~MeV.

\item
Antitriplet and sextet states classified according to their $J^P(nL)$ quantum numbers are shown in Table~\ref{tab:3and6}. The mass difference between $\Xi_c$ and $\Lambda_c$ in the antitriplet states clearly lies between 180 and 200~MeV. Moreover, the mass splitting  between $\Omega_c$ and $\Xi'_c$ is found to be very close to the one between $\Xi'_c$ and $\Sigma_c$ for five different sets of sextet multiplets. This lends a strong support for the quantum number assignment to the sextet states in this work.

\end{itemize}

\section*{Acknowledgments}
CWC thanks the hospitality of Kyoto University for the hospitality during his visit when part of this work was done.  This research was supported in part by the Ministry of Science and Technology of R.O.C. under Grant
Nos.~104-2112-M-001-022 and 104-2628-M-002-014-MY4.

\newcommand{\bi}{\bibitem}

\end{document}